\documentclass[twocolumn]{aastex61}
\usepackage{graphicx}
\usepackage{amsmath}
\usepackage{gensymb}
\usepackage{hyperref}
\usepackage{bm}

\def\DM{{\rm DM}}
\newcommand{\dmu}{${\rm pc \ cm ^ {-3}}$\,}
\def\mJyPerBeam{{\rm mJy\,beam^{-1}}}
\def\JyPerBeam{{\rm Jy\,beam^{-1}}}

\def\degSqrHr{${\rm deg^2\,hr}$\,}
\def\psrHighDM{{\rm PSR J0837$-$4135}\,}
\def\psrLowDM{{\rm PSR J0630$-$2834}\,}

\newcommand{\red}[1]

\shorttitle{Limits on FRB emission}
\shortauthors{Sokolowski et al.}

\begin{document}

\title{No low-frequency emission from extremely bright Fast Radio Bursts}


\correspondingauthor{Marcin Sokolowski}
\email{marcin.sokolowski@curtin.edu.au}

\author[0000-0001-5772-338X]{M. Sokolowski}
\affiliation{International Centre for Radio Astronomy Research, Curtin University, Bentley, WA 6102, Australia}
\affiliation{ARC Centre of Excellence for All-sky Astrophysics (CAASTRO), Australia}

\author[0000-0002-8383-5059]{N.~D.~R. Bhat}
\affiliation{International Centre for Radio Astronomy Research, Curtin University, Bentley, WA 6102, Australia}
\affiliation{ARC Centre of Excellence for All-sky Astrophysics (CAASTRO), Australia}

\author[0000-0001-6763-8234]{J.-P. Macquart}
\affiliation{International Centre for Radio Astronomy Research, Curtin University, Bentley, WA 6102, Australia}
\affiliation{ARC Centre of Excellence for All-sky Astrophysics (CAASTRO), Australia}

\author[0000-0002-7285-6348]{R.~M. Shannon}
\affiliation{International Centre for Radio Astronomy Research, Curtin University, Bentley, WA 6102, Australia}
\affiliation{Australia Telescope National Facility, CSIRO Astronomy and Space Science, PO Box 76, Epping, NSW 1710, Australia}
\affiliation{Centre for Astrophysics and Supercomputing, Swinburne University of Technology, PO Box 218, Hawthorn, VIC 3122, Australia}

\author[0000-0003-2149-0363]{K.~W. Bannister}
\affiliation{Australia Telescope National Facility, CSIRO Astronomy and Space Science, PO Box 76, Epping, NSW 1710, Australia}

\author[0000-0002-3532-9928]{R.~D.Ekers}
\affiliation{Australia Telescope National Facility, CSIRO Astronomy and Space Science, PO Box 76, Epping, NSW 1710, Australia}
\affiliation{International Centre for Radio Astronomy Research, Curtin University, Bentley, WA 6102, Australia}

\author{D.~R.~Scott}
\affiliation{International Centre for Radio Astronomy Research, Curtin University, Bentley, WA 6102, Australia}

\author{A.~P.~Beardsley}
\affiliation{School of Earth and Space Exploration, Arizona State University, Tempe, AZ 85287, USA}

\author{B.~Crosse}
\affiliation{International Centre for Radio Astronomy Research, Curtin University, Bentley, WA 6102, Australia}

\author{D.~Emrich}
\affiliation{International Centre for Radio Astronomy Research, Curtin University, Bentley, WA 6102, Australia}

\author{T.~M.~O.~Franzen}
\affiliation{International Centre for Radio Astronomy Research, Curtin University, Bentley, WA 6102, Australia}

\author[0000-0002-3382-9558]{B.~M.~Gaensler} 
\affiliation{Sydney Institute for Astronomy, School of Physics, The University of Sydney, NSW 2006, Australia}
\affiliation{ARC Centre of Excellence for All Sky Astrophysics in 3 Dimensions (ASTRO 3D), Australia}
\affiliation{Dunlap Institute for Astronomy and Astrophysics, University of Toronto, ON, M5S 3H4, Canada}

\author{L.~Horsley}
\affiliation{International Centre for Radio Astronomy Research, Curtin University, Bentley, WA 6102, Australia}

\author[0000-0003-2756-8301]{M.~Johnston-Hollitt}
\affiliation{International Centre for Radio Astronomy Research, Curtin University, Bentley, WA 6102, Australia}

\author[0000-0001-6295-2881]{D.~L.~Kaplan}
\affiliation{Department of Physics, University of Wisconsin--Milwaukee, Milwaukee, WI 53201, USA}

\author{D.~Kenney}
\affiliation{International Centre for Radio Astronomy Research, Curtin University, Bentley, WA 6102, Australia}

\author{M.~F.~Morales}
\affiliation{Department of Physics, University of Washington, Seattle, WA 98195, USA}
\affiliation{ARC Centre of Excellence for All Sky Astrophysics in 3 Dimensions (ASTRO 3D), Australia}

\author{D.~Pallot}
\affiliation{International Centre for Radio Astronomy Research, University of Western Australia, Crawley 6009, Australia}

\author{G.~Sleap}
\affiliation{International Centre for Radio Astronomy Research, Curtin University, Bentley, WA 6102, Australia}

\author{K.~Steele}
\affiliation{International Centre for Radio Astronomy Research, Curtin University, Bentley, WA 6102, Australia}

\author[0000-0002-8195-7562]{S.~J.~Tingay} 
\affiliation{International Centre for Radio Astronomy Research, Curtin University, Bentley, WA 6102, Australia}

\author[0000-0001-6324-1766]{C.~M.~Trott} 
\affiliation{International Centre for Radio Astronomy Research, Curtin University, Bentley, WA 6102, Australia}
\affiliation{ARC Centre of Excellence for All Sky Astrophysics in 3 Dimensions (ASTRO 3D), Australia}

\author{M.~Walker}
\affiliation{International Centre for Radio Astronomy Research, Curtin University, Bentley, WA 6102, Australia}

\author[0000-0002-6995-4131]{R.~B.~Wayth} 
\affiliation{International Centre for Radio Astronomy Research, Curtin University, Bentley, WA 6102, Australia}
\affiliation{ARC Centre of Excellence for All Sky Astrophysics in 3 Dimensions (ASTRO 3D), Australia}

\author{A.~Williams}
\affiliation{International Centre for Radio Astronomy Research, Curtin University, Bentley, WA 6102, Australia}

\author{C.~Wu} 
\affiliation{International Centre for Radio Astronomy Research, University of Western Australia, Crawley 6009, Australia}

\begin{abstract}
We present the results of a coordinated campaign conducted with the Murchison Widefield Array (MWA) to shadow Fast Radio Bursts (FRBs) detected by the Australian Square Kilometre Array Pathfinder (ASKAP) at 1.4\,GHz, which resulted in simultaneous MWA observations of seven ASKAP FRBs. We de-dispersed the $24\,\times \,1.28$\,MHz MWA images across the $170-200$\,MHz band taken at 0.5\,second time resolution at the known dispersion measures (DMs) and arrival times of the bursts and searched both within the ASKAP error regions (typically $\sim\,10\arcmin\times\,10\arcmin$), and beyond ($4\degree\times4\degree$). We identified no candidates exceeding a $5\sigma$ threshold at these DMs in the dynamic spectra. These limits are inconsistent with the mean fluence scaling of $\alpha=-1.8 \pm 0.3$ (${\cal F}_\nu \propto \nu^\alpha$, where $\nu$ is the observing frequency) that is reported for ASKAP events, most notably for the three high fluence (${\cal F}_{1.4\,{\rm GHz}} \gtrsim 100$\,Jy\,ms) FRBs 171020, 180110 and 180324.
Our limits show that pulse broadening alone cannot explain our non-detections, and that there must be a spectral turnover at frequencies above 200\,MHz. We discuss and constrain parameters of three remaining plausible spectral break mechanisms: free-free absorption, intrinsic spectral turn-over of the radiative processes, and magnification of signals at ASKAP frequencies by caustics or scintillation. If free-free absorption were the cause of the spectral turnover, we constrain the thickness of the absorbing medium in terms of the electron temperature, $T$, to $< 0.03\, (T/10^4\, K)^{-1.35}\,$pc for FRB~171020.
\end{abstract}


\section{Introduction} \label{sec:intro}
The origin of the bright, millisecond-timescale emission associated with Fast Radio Bursts (FRBs) remains an open question. Many of the fundamental observational characteristics of these bursts remain stubbornly elusive, a fact exemplified by poor constraints on even the spectral extent of the radio emission. Until very recent FRB detections by Canadian Hydrogen Intensity Mapping Experiment \citep[CHIME/FRB;][]{2018arXiv180311235T} down to 400\,MHz \citep{2018ATel11901}, the lowest frequency FRB was observed at 700\,MHz \citep{Masuietal2015}. Despite major efforts (including \citet{2014A&A...570A..60C}, \citet{2015MNRAS.452.1254K}, \citet{2015AJ....150..199T}, \citet{2016MNRAS.458.3506R}, \citet{2016Natur.530..453K}, \citet{2017ApJ...844..161A}, \citet{2017ApJ...844..140C} and \citet{2016ApJ...826..223B} to name a few) to date, no FRB emission has been reported below 400\,MHz or above 8\,GHz \citep{2018arXiv180404101G}. Currently the only published limit on the spectral index below 700\,MHz resulting from simultaneous broadband observations is $\alpha < - 3$ (${\cal F}_\nu \propto \nu^{\alpha}$, where ${\cal F}_\nu$ is fluence at the observing frequency $\nu$), for the Parkes FRB~150418 \citep{2016Natur.530..453K}.
Moreover, there have been no coincident detections of FRBs in any other waveband, despite extensive multi-wavelength simultaneous and follow-up observations with optical, IR, X-ray and gamma-ray facilities \citep{Petroffetal2015,scholz16,Lawetal2017,2016Natur.530..453K}.

The dearth of FRB detections at low radio frequencies ($\le 400$\,MHz) presents a critical impediment to the analysis of the burst energetics.  
Burst energies up to $E \sim 10^{35}\,$J are inferred by integrating the emission across the observing band and assuming the emission is isotropic \citep{Lorimeretal2007,Thorntonetal2013,Bannisteretal2017}.  
However, reliable spectral measurements based on the 20 FRBs reported by the Commensal Realtime ASKAP Fast Transients (CRAFT) survey \citep{Macquartetal2010} on the Australian Square Kilometre Array Pathfinder (ASKAP) at 1.4\,GHz, show that the average FRB fluence spectrum is steep, with a spectral index $\alpha = -1.8 \pm 0.3$ \citep[][]{Shannonetal2018}. This indicates that the low-frequency cutoff likely dominates the energetics of the radio emission, notwithstanding the fact that the bursts detected by ASKAP often exhibit patchy spectral structure \citep{Shannonetal2018}, as does the repeating FRB~121102 \citep{Spitleretal2016}. High fluences and steep spectral indices of CRAFT FRBs make them ideal targets for low-frequency observations with the MWA.

Low-frequency measurements also provide diagnostics of the plasma along the line of sight to the FRB.   
In particular, a number of propagation effects potentially influence the spectral characteristics at low frequencies, and their identification in FRB data would place constraints on the burst environment and the properties of the plasma encountered along the ray path.  The most obvious two effects are scattering due to inhomogeneities in the plasma (most likely distant from the source, i.e. $\gg 1\,$pc), which causes temporal smearing of the signal proportional to $\sim \nu^{-4}$\,\citep{2004ApJ...605..759B}, and free-free absorption by dense, circumburst plasma, whose optical depth scales as $\tau_{\rm ff} \sim \nu^{-2.1}$.

The characteristics of the low-frequency emission bear heavily on FRB detection rates with future survey facilities such as the low-frequency component of the Square Kilometre Array  \citep[SKA\_Low;][]{2015aska.confE..51F}. The wide fields-of-view (FoVs) accessible by this telescope render it capable of detecting transient phenomena at extremely high rates \citep{Fenderetal2015}. Moreover, the recent CHIME/FRB detections show that FRBs can be observed at least down to 400\,MHz triggering even more interest in lower frequencies and spectral extent of FRB emission.

In this Letter we present the results of an observing campaign undertaken with the Murchison Widefield Array \citep[MWA;][]{2013PASA...30....7T} to detect low-frequency radio emission coincident with the bright FRBs detected by the CRAFT survey at 1.4\,GHz \citep{Macquartetal2010,Bannisteretal2017,Shannonetal2018}. The MWA shadowed (co-tracked) the pointing positions of the ASKAP antennas, so that the precise dispersion measure (DM), time of arrival and the approximate position (typically a 10\arcmin  $\times$ 10\arcmin~region) of each burst are all known, greatly reducing the searched volume of parameter space relative to a blind survey.


\section{Observations and Data Reduction} \label{sec:obsn}

\subsection{Observing Strategy}
\label{sec:mwa_data}

\begin{figure}
  \begin{center}
  \includegraphics[width=3.4in]{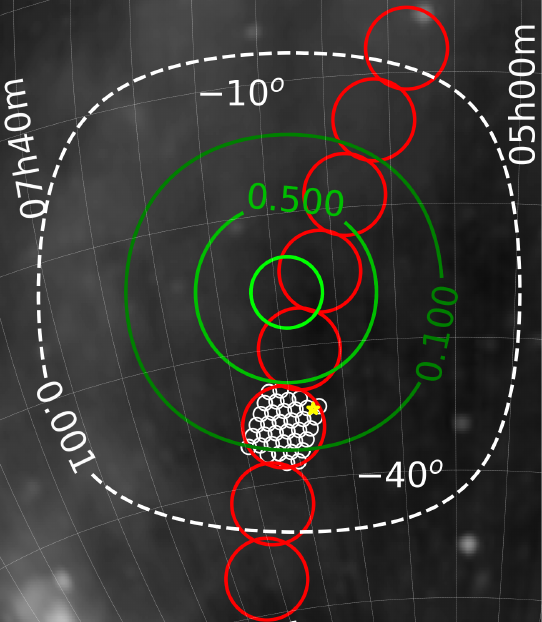}  
  \caption{An illustration of the MWA shadowing strategy for FRB~180324 (on top of the image by \citet{1982A&AS...47....1H} at 408\,MHz). The large ($\approx 7\degree$) red circles represent coverage of ASKAP antenna beams distributed near Galactic latitude -20$\degree$ and small white circles ($\approx 1.2\degree$) are the individual 36 beams of ASKAP antenna 25, which detected FRB~180324 (yellow star). The green (at 0.5 and 0.1) and white-dashed (at 0.001 with angular size $\approx 42\degree\times42\degree$) contours show the MWA beam \citep{2017PASA...34...62S} normalized to maximum response at zenith.}
  \label{fig_frb180324_strategy}
  \end{center}
\end{figure}

The observations were made with the MWA whilst the ASKAP antennas observed in the fly's-eye mode\footnote{Individual antennas pointing in different directions and covering $\approx$300\,deg$^{2}$ (with 10 ASKAP antennas).} distributed along a certain Galactic latitude \citep[see][]{Bannisteretal2017}. The FoV of the MWA in the frequency band $170-200$\,MHz is approximately 450\,deg$^{2}$ (FWHM$\sim 21\degree$), enabling nearly full coverage of the ASKAP fly's eye FoV with sensitivity $\gtrsim 50$\% of the primary beam. Moreover, the selected frequency band minimizes the effects of pulse broadening (due to scattering) and radio-frequency interference. Figure~\ref{fig_frb180324_strategy} shows the observing setup during FRB~180324.

Table~\ref{tab_frbs} summarizes the ASKAP FRBs \citep[submitted]{Shannonetal2018,JP2018} detected while MWA shadowed. 
The strategy succeeded for the first time with FRB~171020, and the MWA collected data before, during and after the FRB detection by ASKAP (Table~\ref{tab_frbs}).
The MWA data were collected during the transition to the extended array \citep{RW2018_mwa_phase2}, implying reduced sensitivity (only $\sim$70\% of antennas). Since then, six more ASKAP FRB positions were observed by the MWA in a similar mode. In all cases MWA data were recorded in 10-kHz frequency and 0.5-second temporal resolutions.

The shadowing program was performed mostly during daytime. Hence, only FRBs~171020 and 180324 were detected by ASKAP after sunset whilst the FRB~180110 field was observed with the Sun close to a ``null'' of the MWA's primary beam (a direction with very low sensitivity). In the other cases the data quality was too low to derive meaningful limits due to the presence of the Sun in the sidelobe of the primary beam. 

\subsection{Calibration}
\label{sec:calibation}

The MWA data were calibrated with the Common Astronomy Software Applications \citep[CASA;][]{casa}, using observations of a calibrator source (3C444 or Pictor A). 
We applied calibration solutions to FRB field observations and created dirty images in 0.5-second temporal (shortest possible) and 1.28-MHz frequency resolutions using \textsc{wsclean} \citep{wsclean} and natural weighting.

We also calibrated and imaged MWA observations (292\,seconds) collected before the FRB observations in order to create reference images of the FRB fields (typical standard deviation of noise $\sigma \sim 20-40$~$\mJyPerBeam$). The flux density scale was calibrated using sources from the GaLactic Extragalactic All-sky MWA (GLEAM) survey \citep{gleam_nhw} identified in the reference images. 
\begin{table*}
\caption {The details of the ASKAP FRBs shadowed by the MWA}
\begin{center}
\begin{tabular}{@{}cccccccccccc@{}}
\hline\hline
    &               &                   &                         &               &                 &  &                                    & \multicolumn{3}{c}{    $\mathcal{F}_{\rm \textsc{185MHz}}$ [Jy\,ms] expected$^f$ } &  \\[-5pt]
FRB & UTC           &$\rm DM_{tot}$$^a$ &  $\rm DM_{mw}$$^a$      & $t_{arr}$$^b$ & $t_{sweep}$$^c$ & $\tau_{scat}$$^d$ & $\mathcal{F}_{\rm \textsc{1.4GHz}}^e$ &  $\alpha=-1$ & $\alpha=-2$ & $\alpha = -1.8^g$ &  $\mathcal{U}_{5\sigma}$$^h$ \\[-5pt]
    & detection     &\multicolumn{2}{c}{$[\mathrm{pc/cm^{3}}\,]$} &     [s]       & [s]             & [ms] & [Jy\,ms\,]                     &              &             &                                         &  $[\mathrm{Jy\,ms}\,]$\\[-5pt] 
    &               &                   &                         &               &                 &     &                                &              &             &                                         &  \\[-5pt]
    &               &                   &                         &               &                 &     &                                &              &             &                                         &   \\
\hline%
171020        &  10:27:59.00 &  114.1 & 38.4 & 11.7  & 4.5   & 1.7 & $200^{+500}_{-100}$ & $1500^{+4000}_{-800}$ & $11400^{+30000}_{-6000}$ & $7600^{+19000}_{-4000}$ & 2200 \\ 
180110        &  07:34:34.95 &  715.7 & 38.8 & 73.0  & 28.0  & 4.5 & $420^{+20}_{-20}$   & $3200^{+150}_{-150}$ & $23900^{+1100}_{-1100}$  & $16000^{+800}_{-800}$ & 3350$^i$ \\[-8pt] 
          &              &        &   &    &       &         &     &                 &                      &                       & or  \\[-7pt]
          &              &        &   &    &       &         &     &                 &                      &                       & 6500$^j$ \\
180128.0  &  00:59:37.97 &  441.4 &  31.5  & 45.0  & 17.3    & 2.9 & $51^{+2}_{-2}$  & $380^{+15}_{-15}$ & $2900^{+110}_{-110}$ & $1940^{+80}_{-80}$   & GL$^k$ \\
180128.2  &  04:53:26.80 &  495.9 &  41.0  & 50.6  & 19.40   & 2.3 & $66^{+4}_{-4}$  & $500^{+30}_{-30}$ & $3800^{+230}_{-230}$ & $2500^{+150}_{-150}$   & SL$^k$ \\
180130    &  04:55:29.99 &  343.5 &  39.0  & 34.90 & 13.35   & 6.0 & $95^{+3}_{-3}$  & $720^{+20}_{-20}$ & $5400^{+170}_{-170}$ & $3600^{+110}_{-110}$   & SL$^k$ \\
180315    &  05:05:30.99 &  479.0 &  101.7 & 48.66 & 18.63   & 2.4 & $56^{+4}_{-4}$  & $420^{+30}_{-30}$ & $3200^{+230}_{-230}$ & $2100^{+150}_{-150}$ & SL$^k$ \\
180324    &  09:31:46.70 &  431.0 &   64.0 & 43.79 & 16.75   & 4.3 & $71^{+3}_{-3}$  & $540^{+20}_{-20}$ & $4000^{+170}_{-170}$ & $2700^{+110}_{-110}$ & 450$^i$ \\ 
\hline\hline
\end{tabular}
\end{center}
\tablenotetext{a}{$\rm DM_{tot}$ : the total DM measured by ASKAP; $\rm DM_{mw}$ is the contribution of the Milky Way (from NE2001; \citet{NE2001})}
\tablenotetext{b}{$t_{arr}$ : the time delay between ASKAP detection at 1297\,MHz, and the expected arrival at 200\,MHz}
\tablenotetext{c}{$t_{sweep}$ : the sweep time over the MWA observing band ($170-200$\,MHz)}
\tablenotetext{d}{$\tau_{scat}$ : scattering time at 1.4\,GHz fitted to ASKAP data for 180110 and 180130 and pulse width for other FRBs.} 
\tablenotetext{e}{The errors represent 90\% confidence limits}
\tablenotetext{f}{The fluences are extrapolated to 185.6\,MHz assuming power law scaling ${\cal F}_\nu \propto \nu^{\alpha}$}
\tablenotetext{g}{The mean spectral index of ASKAP FRBs.} 
\tablenotetext{h}{$\mathcal{U}_{5\sigma}$ : the MWA $5\sigma$ upper limit on $\mathcal{F}_{\rm \textsc{185MHz}}$}
\tablenotetext{i}{These limits are higher by a factor $\approx1.2-1.5$ because of de-dispersion in 1.28\,MHz channels}
\tablenotetext{j}{This FRB was significantly temporally broadened (see Sec.~\ref{subsec:frbalgo}); therefore, we also present the limit from de-dispersed images averaged over 10\,s (Tab.~\ref{tab_mwa_upper_limits})}
\tablenotetext{k}{The data quality was too low to obtain meaningful upper limits due to the presence of the Sun in sidelobe (SL) or grating-lobe (GL)}
\label{tab_frbs}
\end{table*}

\subsection{FRB detection algorithm}
\label{subsec:frbalgo}

\begin{figure*}
  \begin{center}
  \includegraphics[width=\textwidth]{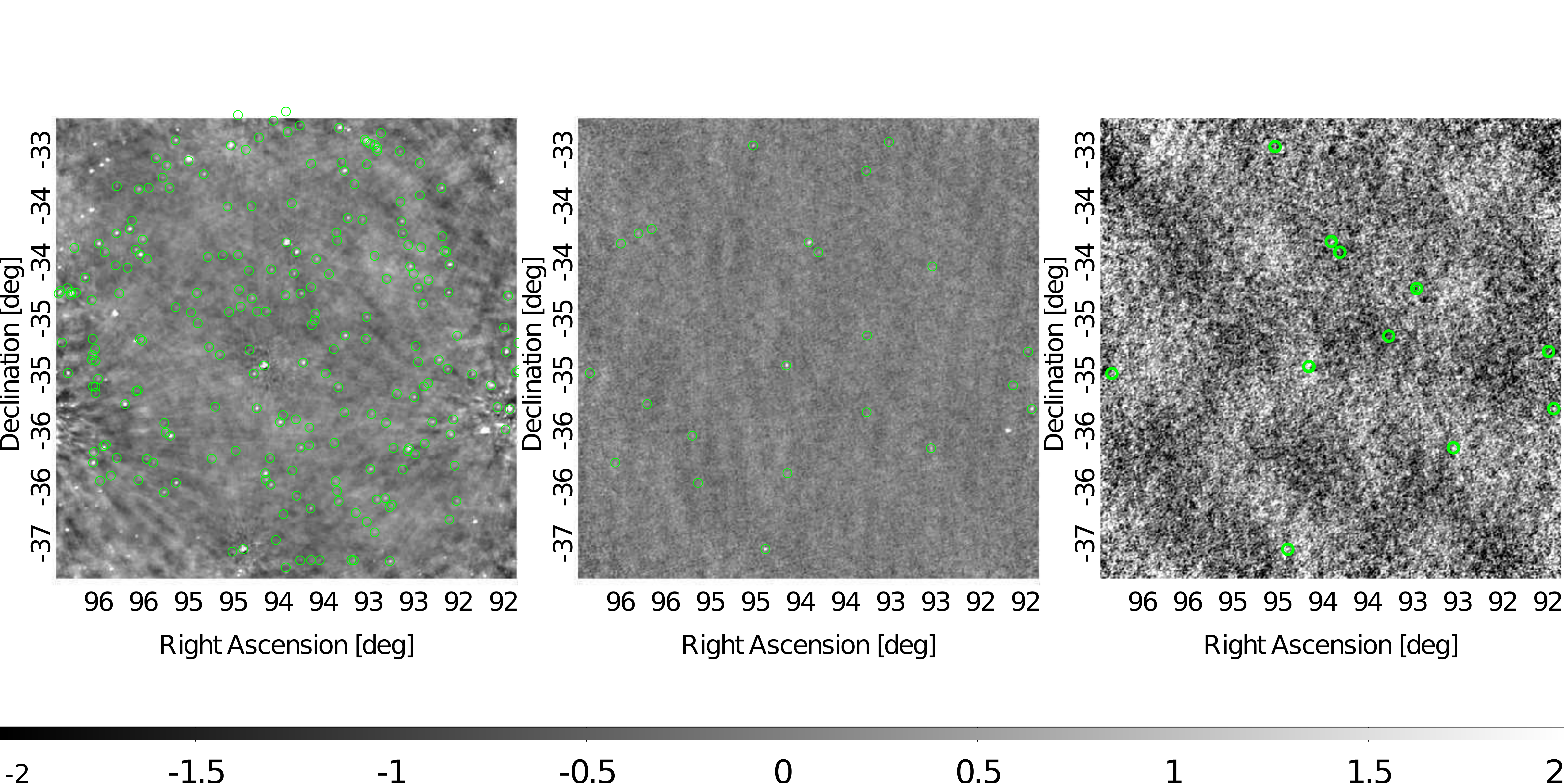}
  \caption{Example images of the FRB~180324 position (image centers). (Left) The reference image of the field (the mean of 6 observations without the FRB), with noise in the center $\sigma\approx$~17\,$\mJyPerBeam$; the green circles are NVSS \citep{nvss} sources brighter than 10\,mJy at 1.4\,GHz. The image scale spans $-0.2-0.2$\,$\JyPerBeam$. (Center) An example 0.5-s dirty image obtained by stacking the appropriately time-shifted $24\,\times\,1.28$\,MHz spectral channels (i.e. de-dispersing) to \DM=431.0\,\dmu; the noise in the image center is $\sigma\approx$~180\,$\mJyPerBeam$ and the green circles are NVSS sources brighter than 100\,mJy at 1.4\,GHz. (Right) An example single 0.5-s dirty image in a single 1.28\,MHz channel; noise in the image center is $\sigma\approx$~760\,$\mJyPerBeam$ and the green circles are NVSS sources brighter than 200\,mJy at 1.4\,GHz. The scales in the center and right images are as shown in the colorbar (in $\JyPerBeam$).}
  \label{fig_frb180324}
  \end{center}
\end{figure*}

The transient detection algorithm used the reference images (Sec.~\ref{sec:calibation}) to create a list of reference sources above certain threshold ($5\sigma$). The 24$\times $1.28\,MHz, 0.5-second images across the observing band were time-aligned according to the ASKAP DM and summed to provide a 0.5-second resolution de-dispersed image (example images are shown in Fig.~\ref{fig_frb180324}). We also examined the de-dispersed images before and after the expected burst arrival time (we inspected all 0.5-s images within the analyzed MWA observation). The details of candidates exceeding a $5\sigma$ threshold, which were not present in the list of reference sources, were saved for further visual inspection. We did not observe any $>5\sigma$ event within the ASKAP error boxes\footnote{The expected number of candidates exceeding $5\sigma$ due to Gaussian noise fluctuations is $\ll$1 for $\sim$100$\times$100 pixels.}.

The 5$\sigma$ transient candidates identified within a larger (4\degree$\times$4\degree) field were visually inspected on de-dispersed images and dynamic spectra of the candidate pixels, and none of them showed any signs of dispersion sweep in the dynamic spectra. As a final check, we visually inspected all 0.5-second, 1.28-MHz images.

FRB~180110 was significantly scattered (Tab.~\ref{tab_frbs}), with an expected pulse width at the MWA's frequency, $\tau_{\rm\textsc{MWA}}\approx$5.4 seconds (Sec.~\ref{sec:Discussion}). Therefore, in the second part of the algorithm we averaged over multiple de-dispersed (and non-dedispersed) images on timescales of 5, 10, 15, 20, 25 and 30 seconds. Yet, we did not identify any transient events exceeding 5$\sigma$ threshold with any signs of dispersion sweep in the dynamic spectra of the candidate pixels. These time-averaged images were also visually inspected. The algorithm was executed on Stokes I and V images because noise in 0.5-second V images was slightly lower ($\sim 20-30$\%). The above procedure resulted in upper limits on the flux densities and consequently fluences of low-frequency counterparts of three ASKAP FRBs, which are summarized in Tables~\ref{tab_frbs} and ~\ref{tab_mwa_upper_limits}.

\subsection{Verification of the FRB Pipeline}
\label{sec_pipeline_verification}

\begin{figure*}
  \begin{center}
  \includegraphics[width=4in]{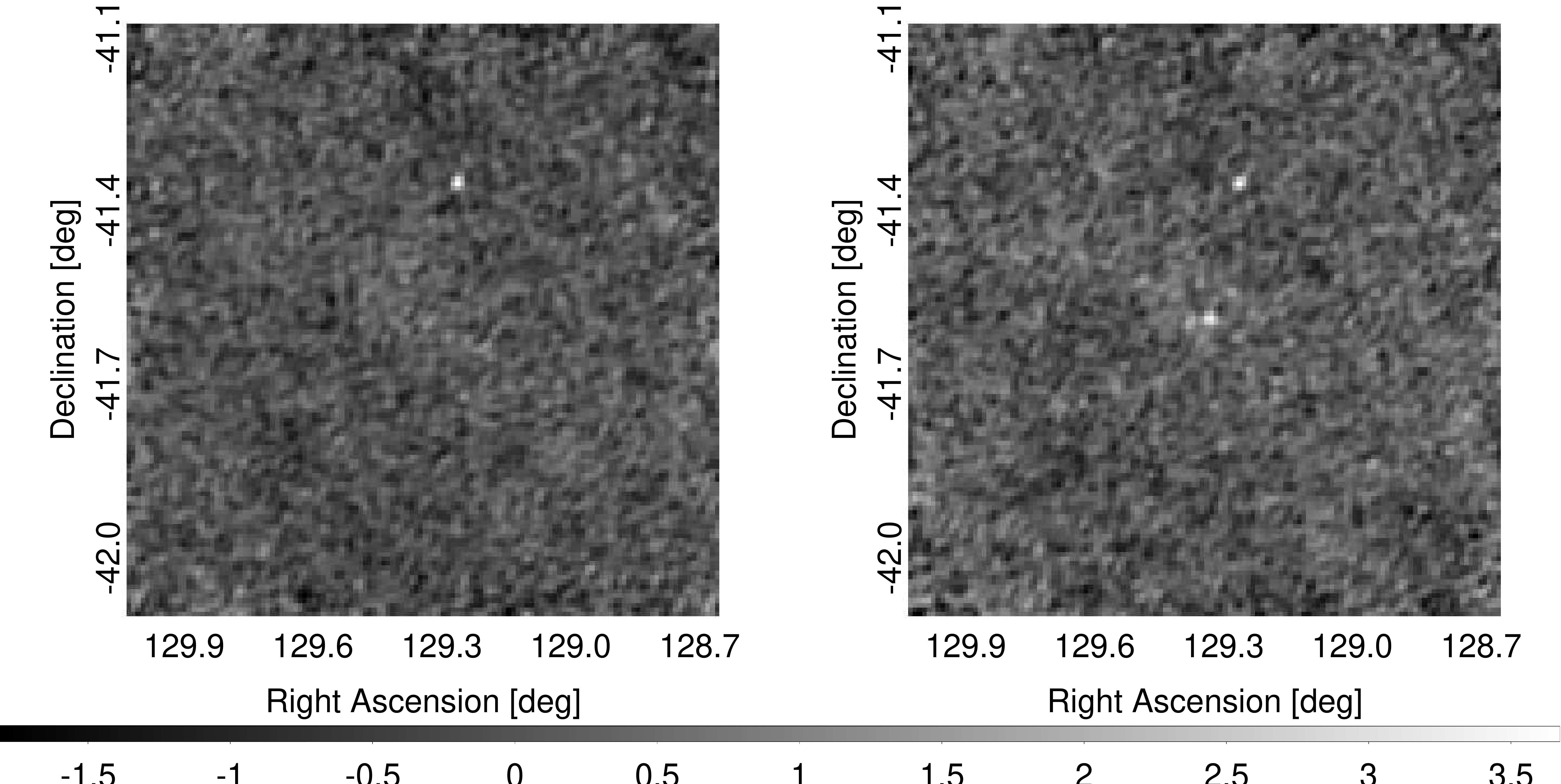}
  \includegraphics[width=3in]{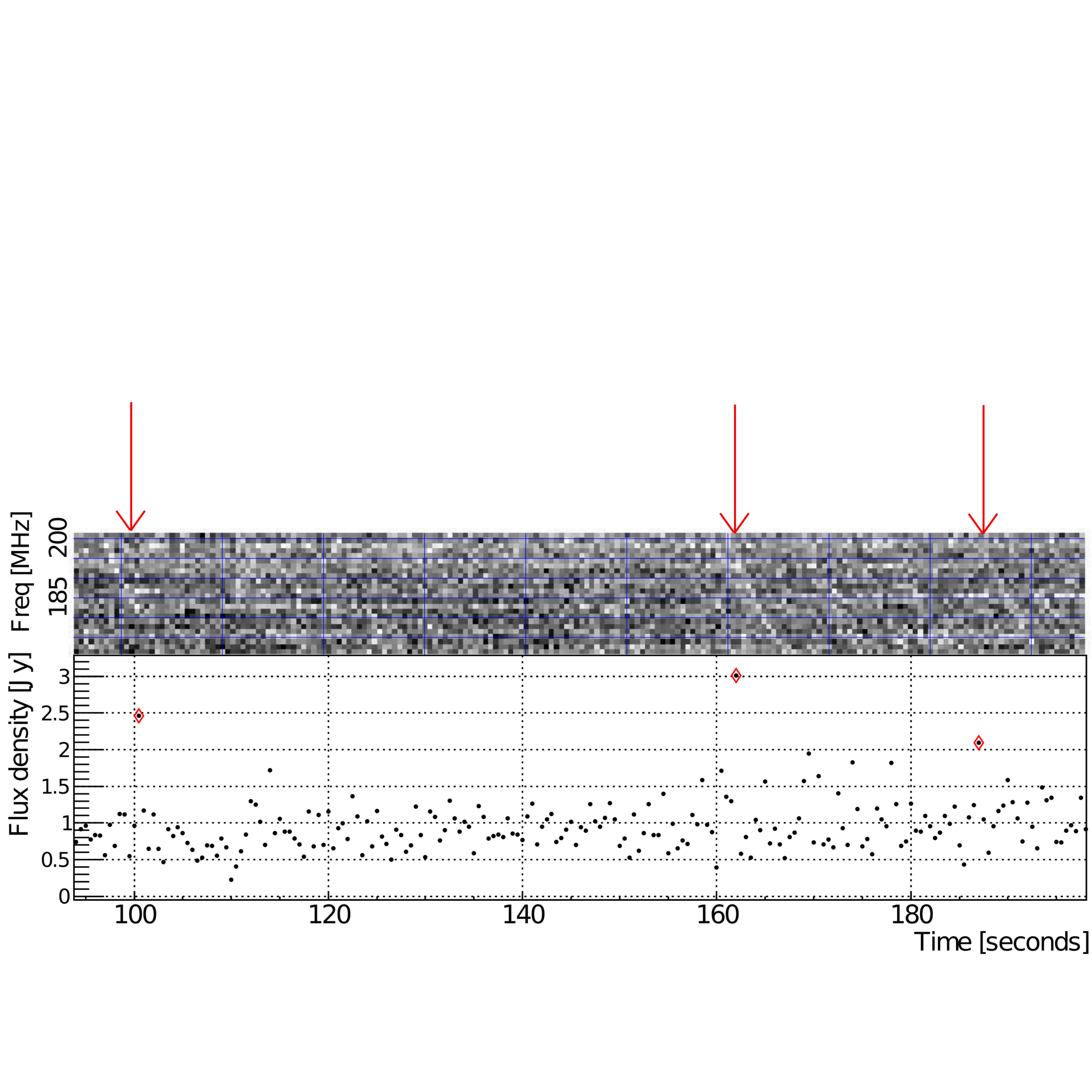} 
  \caption{An example of de-dispersed images of PSR J0837-4135 (left) without a pulse, and (center) with a pulse detected by the algorithm. (Right) the corresponding dynamic spectrum of the central pixel showing the frequency sweep due to DM=147.29\,\dmu (right upper image). The three sweeps (sweep time $\approx$5.8\,seconds in the $170-200$\,MHz range) can be seen at times ranges $97-105$, $160-166$ and $183-191$\,seconds between the red arrows and the data points surrounded by the red diamonds marking the three brightest points in the lightcurve from the de-dispersed images (right lower image).}
  \label{fig_0837}
  \end{center}
\end{figure*}

The algorithm was verified using a relatively high-DM pulsar (147.29\,\dmu), \psrHighDM, with a pulse period P$\approx$0.751\,s \citep{2005AJ....129.1993M}.
The pipeline detected three individual bright pulses above 5$\sigma$ in a 112-second ($\approx$ 149 pulses) observation. Example images with and without a pulse, the dynamic spectrum and the lightcurve observed in the de-dispersed images are shown in Figure~\ref{fig_0837}.

The efficiency of the algorithm was verified on a 112-second observation of a lower-DM pulsar (34.425\,\dmu), \psrLowDM, with P$\approx$1.244\,s \citep{2005AJ....129.1993M}. In the 112-second (90 pulses) observation, the algorithm identified 31 pulses above the $5\sigma$ threshold in the de-dispersed images (110 pulses with the $3\sigma$ threshold). Due to lack of pulsars with a suitable combination of DM and scattering time, we have verified the algorithm on averaged images by ``injecting'' a simulated FRB signal into 0.5-second MWA images and confirming that it was detected in the averaged images.
These tests confirmed that the algorithm is capable of detecting highly-dispersed, FRB-like transients in single and averaged 0.5-s MWA images.

%

\section{Discussion} 
\label{sec:Discussion}

%
%
%
%
%

The absence of low-frequency emission coincident with any of the shadowed CRAFT FRBs places constraints on the emission characteristics of these bursts and their environments.  We provide a brief discussion here in the context of the four most obvious interpretations.

%
(i) Temporal smearing due to scattering through an inhomogeneous plasma potentially dominates over the intrinsic pulse width at low frequencies and decreases the burst detectability.  An upper limit to the low-frequency pulse width is inferred by attributing the ASKAP-derived pulse width entirely to scattering, and assuming that pulse broadening scales as $\nu^{-3.5}$ (or $\nu^{-4}$)\footnote{\citet{Shannonetal2018} measure $\tau_{\rm scat} \propto \nu^{-3.5 \pm 0.5}$ in the ASKAP band, so we examine the consequences of both a $\nu^{-3.5}$ and a $\nu^{-4}$ scaling.}. The measured pulse widths of FRBs 171020, 180110 and 180324 at 1400\,MHz in turn imply limits on the pulse width at 185\,MHz of 2.1 (5.6), 5.4 (14.8) and 5.1 (14) \,seconds respectively.  Since pulse fluence is unaltered by scattering, the fluence upper limits relative to the expected fluence (Tables~\ref{tab_frbs} and \ref{tab_mwa_upper_limits}) show that any pulse with a spectrum steeper than $\alpha \approx -1$ would have been detected.  We conclude that pulse broadening {\it alone} is insufficient to explain the MWA non-detections because the spectral index of the emission would have to flatten substantially between 185\,MHz and 1295\,MHz from the mean value of $\alpha = -1.8$ measured for the ASKAP FRB population \citep{Shannonetal2018}. 

(ii) The viability of the hypothesis that free-free absorption is responsible for the absence of low frequency emission may be cast in terms of a limit on the thickness of the absorbing material. An upper limit to the depth of the absorbing region is derived by assuming that all plasma not attributable to the Milky Way is confined to the absorbing medium (not to the intergalactic medium, thereby assuming the redshift term (1+z) in Eq.~\ref{eq_delta_l_limit} is $\approx 1$), whose emission measure is then ${\rm EM} = \int N_e^2 dl = {\rm DM}_{\rm extra}^2 \,\Delta L^{-1}\,$pc\,cm$^{-6}$, where $\Delta L$ is the region thickness in pc, $N_e$ is electron density, and ${\rm DM}_{\rm extra}$ is the DM excess (in \dmu).  A burst whose predicted flux density is $S_{\rm pred}$ but is undetected at a limit of $S_{\rm lim}$ constrains the free-free optical depth to $\tau_{\rm ff} > \ln (S_{\rm pred}/S_{\rm lim})$, and one has
\begin{eqnarray}
1.1 \times 10^{-5} \, T_{4}^{-1.35} \nu_{185}^{-2.1} \left( \frac{{\rm DM}_{\rm extra}^2}{\Delta L}\right) > \ln \left( \frac{S_{\rm pred}}{S_{\rm lim}} \right),
\label{eq_delta_l_limit}
\end{eqnarray}
where $T_{4}$ is the electron temperature in units of $10^4\,$K and $\nu_{185}$ is the observing frequency, normalized to 185\,MHz. The redshift term is not included as the equation still leads to a valid upper limit. We assume the filling factor $f=1$, and if $f<1$ the limit is $\Delta L \propto f^{-1}$. The DM values in excess of the Milky Way contributions predicted by NE2001 \citep{NE2001} are 75.7, 676.9 and 367.0\,\dmu for FRBs 171020, 180110 and 180324 respectively.  We conservatively subtract a further 15\,pc\,cm$^{-3}$ for the Milky Way halo contribution \citep[see the discussion in][]{Shannonetal2018}.  For a burst spectral index of $\alpha=-1.8$, and using $5\sigma$ limits for $S_{\rm lim}$ derived from Table \ref{tab_mwa_upper_limits}, this implies upper limits on $\Delta L$ of $3 \times 10^{-2} \,T_4^{-1.35}\,$pc, $5 \,T_4^{-1.35}\,$pc (using ${\cal U}_{5\sigma} = 6500\,$Jy\,ms) and $0.7 \, T_4^{-1.35}\,$pc for these three bursts respectively.  It is not possible to place significant constraint on $\tau_{\rm ff}$ for $\alpha\,>\,-1$. The tightest constraint on $\tau_{\rm ff}$ from FRB~171020 is likely tighter because a fraction of the electron column density is associated with host galaxy interstellar medium (ISM), intergalactic medium (IGM), or both.


\begin{table}
\begin{center}
\begin{tabular}{@{}ccccccc@{}}
\hline\hline
 FRB  & $\sigma_{dd}$ & $\sigma_{0.5s}$ & $\sigma_{5s}$ & $\sigma_{10s}$ & $\sigma_{15s}$ & $\sigma_{20s}$ \\
\hline
20171020 & 880 & 4570 & 400 & 350 & 330 & 310         \\ 
20180110 & 1340 & 6640 & 700  & 650  & 620  & 600     \\ 
20180324 & 180 & 840 & 120  & 120  & 110  & 110       \\ 
\hline\hline 
\end{tabular}
\end{center}
\caption{MWA 1$\sigma$ noise of flux density ($\mJyPerBeam$) in the centers of beam-corrected images on average of 24 de-dispersed 1.28-MHz channels ($\sigma_{dd}$), single 0.5-second images ($\sigma_{0.5s}$) and averaged over 5 ($\sigma_{5s}$), 10 ($\sigma_{10s}$), 20 ($\sigma_{20s}$) and 30\,seconds ($\sigma_{30s}$).}
\label{tab_mwa_upper_limits}
\end{table}

(iii) The flux densities of the bursts detected by ASKAP could be enhanced by plasma lensing effects due to caustics or scintillation \citep{Cordesetal2017}.  It is improbable that any lensing effects, if present, extend over an order of magnitude in frequency, so it is unlikely that it plays a significant role both at ASKAP and low frequencies. 
The ratios of the measured ASKAP fluences to their 185-MHz upper limits yield a lower limit to the magnification at 1.4\,GHz. Comparing against the 5$\sigma$ upper limit at $\alpha =-1.8$, we find magnifications $\mu > 3, 2$ \& $6$ for FRBs 171020, 180110 and 180324 respectively.

Most ASKAP FRBs exhibit fine frequency structure, some with a factor $>10$ deviations from the mean, with a filling factor of $\sim\,5\%$ \citep{Shannonetal2018}. If this structure is due to diffractive scintillation, the decorrelation bandwidth at 185\,MHz would be $\sim\,10^{-3}$ smaller than that at 1.4\,GHz (assuming $\sim \nu^{4}$ scaling), the spectrum would therefore appear spectrally smooth at low frequency at our resolution, and the low spectral occupancy observed at 1.4\,GHz would not explain the MWA non-detections. However, even if the low spectral occupancy is an intrinsic feature of the emission, it is implausible to attribute the low-frequency non-detections for all three well-constrained FRBs to this effect alone, since the relative spectral bandwidths of ASKAP and the MWA are comparable at $\Delta\,\nu/\nu$ equal to 26\% and 17\% respectively.

(iv) A break in the intrinsic spectrum is a natural explanation for the non-detection and would be plausible on two grounds: (a) for a spectrum with $\alpha\,<\,-1$ the finite burst energy is dominated by the low-energy cutoff, and (b) many pulsars, whose emission resembles the properties of FRBs in some respects, also exhibit spectral breaks in the region $100-300\,$MHz (see Fig.\,7 of \citet{Bilousetal2016} or Fig.\,6 of \citet{2017PASA...34...20M} for a recent summary).
The limit on the spectral cutoff of $\nu_{\rm lo}\,>\,200\,$MHz for $\alpha=-1.8$ implies that the total burst energy cannot exceed 18 times more than the pulse energy measured within the ASKAP band alone.

The presented simultaneous MWA observations of extremely bright ASKAP FRB are a substantial advance over previous surveys conducted below 200\,MHz. Therefore, the derived limits enable us to place tighter constraints on the physics.  
Indeed, no previous low-frequency survey has a sufficiently large exposure to have detected the low-frequency counterpart to the FRBs detected by CRAFT: the $37 \pm 8\,$events\,day$^{-1}$\,sky$^{-1}$ burst detection rate at ${\cal F}_{1.4\,{\rm GHz}} > 26\,$Jy\,ms  
is equivalent to one event every $27_{-5}^{+7}\,\times\,10^3\,$\degSqrHr \citep{Shannonetal2018}. This substantially exceeds the $14\,\times\,10^3\,$\degSqrHr exposure of the LOFAR Pilot Pulsar Survey at 140\,MHz \citep{2014A&A...570A..60C} with a fluence cutoff ${\cal F_{\rm c}} = 2.75 \times 10^3 \Delta T^{1/2}\,{\rm Jy\,ms},$ for pulse widths $\Delta T \leq 1.26\,$s in the range \DM=2-3000\,\dmu. 
The ARTEMIS survey at 150\,MHz \citep{2015MNRAS.452.1254K} examined ${\rm DM} \le 310$\,\dmu so would have missed 6 of the 7 FRBs reported here. Moreover, it only searched for pulse durations below 21\,ms, with a $10\sigma$ threshold ${\cal F_{\rm c}} = 4.47 \times 10^3 \Delta T^{1/2}\,{\rm Jy\,ms}$, where $\Delta T\,$ is burst width in seconds.

The blind pilot survey of \cite{2015AJ....150..199T} had an exposure of only $4.2\,\times\,10^3\,$\degSqrHr with a $7\sigma$ detection limit in a de-dispersed time series of 350\,$\mJyPerBeam$ on a timescale of 2\,s (i.e. a limiting fluence of $700\,$Jy\,ms).  The survey of \cite{2016MNRAS.458.3506R} was sensitive to an event rate a factor of 8.5 lower than that of \cite{2015AJ....150..199T} but only resolved down to a time resolution of 28\,s with a limiting fluence of $7980\,$Jy\,ms.

Our results exceed limits from previous MWA observations.   The $3\sigma$ upper limit on the fluence of FRB~150418 at 185\,MHz, based on shadowing of the Parkes radio telescope, is 1050\,Jy\,ms \citep{2018MNRAS.473..116K}. However, the Parkes burst fluence of $2.0_{-0.8}^{+1.2}\,$Jy\,ms permits only a poor constraint on the spectral index between 1382\,MHz and 185\,MHz of $\alpha > -3$. 

\section{Conclusions} \label{sec:Conclusions}

We have used the MWA to attempt the co-detection of the low-frequency counterparts of the high fluence FRBs detected by the CRAFT survey. None were found. The shadowing strategy employed here enabled us to significantly reduce the searched FRB parameter volume in space, DM and time.  We are able to derive strong upper limits based on the rare, very high ASKAP fluence (${\cal F}_{1.4\,{\rm GHz}}\,\gtrsim$\,26\,Jy\,ms) FRBs, which would require ``blind'' surveys of enormous exposures ($\sim 27000$\,deg$^2\,$h per event) to obtain comparable limits.  

No low-frequency counterparts of these FRBs were identified. The fluences of CRAFT FRBs are accurately measured, and the presented limits on broadband spectral indexes ($\alpha\,\gtrsim\,-1$) supersede the only direct broadband limit existing in the literature ($\alpha\,>\,-3$) obtained with the MWA and Parkes FRB~150418. These limits are strongly at variance with the mean $\alpha =-1.8 \pm 0.3$ spectral index for these FRBs measured within the ASKAP band.

We show that pulse broadening alone due to scattering is insufficient to explain the non-detections of the shadowed FRBs. We place limits on the thickness of a dense medium close to the progenitors under the assumption that free-free absorption is responsible for the spectral turnover; the tightest constraint is for FRB~171020, for which the medium thickness is $<0.03\,T_4^{-1.35}$\,pc. In this case the implied absorption region is so small that it casts considerable doubt that free-free absorption is responsible for the spectral turnover. The speed of the dense wind of any possible massive progenitor, or the shock associated with a young (older than 1 minute) SN would easily encompass a larger region than that implied here. Moreover, \citet[submitted]{Mahony2018} examines the properties of the likely host galaxy of this FRB, and finds no obvious bright compact knots of continuum radio emission at frequencies up to 21\,GHz down to 31.8\,$\mu Jy$. 
If FRBs are enhanced by caustics in the ASKAP band, we constrain the magnification (between 1295 and 185\,MHz) to be $\mu > 6$, in the case of FRB~180324. The largely unconstrained properties of the host galaxy ISM and intervening IGM make it difficult to assess the viability of these implied lensing magnifications. However, the analysis in \citet[submitted]{JP2018} shows that the spectral structure of the bursts at 1.4\,GHz appears to be qualitatively inconsistent with the effect of plasma lensing, which disfavors lensing as the explanation of the MWA non-detections. A turn-over in the intrinsic FRB spectrum, however, remains a plausible explanation. 

For $\alpha \leq -1$ the low-frequency cutoff dominates the burst energy. Thus, for $\alpha =-1.8$ our non-detections limit the total burst energy to $<18$ times the $\sim 10^{35}\,$J values deduced for FRB detections in the ASKAP and Parkes bands alone.

To date, no FRB has been detected simultaneously at high and low frequencies. The on-going enhancements to the MWA will improve the detection sensitivity for FRB searches by about an order of magnitude (owing to much higher temporal and spectral resolutions possible via the voltage buffer mode). These improvements will significantly increase the prospects of detecting low-frequency emission if it is much fainter than our currently set limits, which may warrant a review of FRB models and physics.



\acknowledgments

This scientific work makes use of the Murchison Radio-astronomy Observatory (MRO), operated by CSIRO. We acknowledge the Wajarri Yamatji people as the traditional owners of the Observatory site. Support for the operation of the MWA is provided by the Australian Government (NCRIS), under a contract to Curtin University administered by Astronomy Australia Limited. The ASKAP is part of the Australia Telescope National Facility which is managed by CSIRO. Operation of ASKAP is funded by the Australian Government with support from the National Collaborative Research Infrastructure Strategy. ASKAP uses the resources of the Pawsey Supercomputing Centre. Establishment of ASKAP, the MRO and the Pawsey Supercomputing Centre are initiatives of the Australian Government, with support from the Government of Western Australia and the Science and Industry Endowment Fund. JPM, RMS \& KWB acknowledge Australian Research Council grant DP180100857. R.M.S.  acknowledges salary support through Australian Research Council (ARC) grants FL150100148 and CE170100004. 
Parts of this research were supported by the Australian Research Council Centre of Excellence for All Sky Astrophysics in 3 Dimensions (ASTRO 3D), through project number CE170100013. Parts of this research were conducted by the Australian Research Council Centre of Excellence for All-sky Astrophysics (CAASTRO), through project number CE110001020.


\bibliographystyle{aasjournal}
\bibliography{references-CRAFT_Shadow}


\end{document}